\documentclass[aps,pra,amsmath,amssymb,twocolumn,groupedaddress,showpacs,showkeys]{revtex4-1}
\usepackage{graphicx}  
\usepackage{dcolumn}   
\usepackage{bm}        
\usepackage{amssymb}   
\usepackage{amsmath}
\usepackage{float}
\usepackage{color}
\usepackage{physics}
\usepackage{latexsym}
\usepackage{gensymb}
\usepackage{xcolor}
\usepackage{textcomp}

\usepackage{etoolbox} 
\usepackage{lipsum} 
\usepackage[capitalize]{cleveref}

\makeatletter
\appto{\appendix}{%
  \@ifstar{\def\theequation@prefix{A.}}%
          {}%
}
\makeatother

\begin{document}



\title{Canceling the cavity length induced phase noise in an optical ring cavity for phase shift measurement and spin squeezing}

\author{Enlong Wang}
\author{Gunjan Verma}
\author{Jonathan N. Tinsley}
\author{Nicola Poli}
\email{Also at CNR-INO.}
\author{Leonardo Salvi~~}
\email{Email: leonardo.salvi@unifi.it}


\affiliation{Dipartimento di Fisica e Astronomia and LENS -
Universit\`{a} di Firenze, INFN - Sezione di Firenze, Via Sansone
1, 50019 Sesto Fiorentino, Italy}

\date{\today}

\begin{abstract}

We demonstrate a new method of light phase shift measurement using a high-finesse optical ring cavity which exhibits reduced phase noise due to cavity length fluctuations. 
Two laser beams with a frequency difference of one cavity free spectral range are simultaneously resonant with the cavity, demonstrating noise correlations in the error signals due to the common-mode cavity length fluctuations. The differential error signal shows a 30~dB reduction in cavity noise down to the noise floor in a frequency range up to half the cavity linewidth ($\delta\nu/2 \simeq 30$~kHz). Various noise sources are analyzed and their contributions to the noise floor are evaluated.
Additionally, we apply this noise-reduced phase shift measurement scheme in a simulated spin-squeezing experiment where we have achieved a factor of 40 improvement in phase sensitivity with a phase resolution of 0.7~mrad, which may remove one important barrier against attaining highly spin-squeezed states. 
The demonstrated method is the first reported measurement using an optical ring cavity and two independent beams, a flexible situation. 
This method can find direct application to non-destructive measurements in quantum systems, such as for the generation of spin-squeezed states in atom interferometers and atomic clocks.



\end{abstract}


\maketitle


\section{Introduction}
One of the main limitations in monitoring a quantum system lies in the destruction of the quantum states when a measurement is performed.
In recent years, non-destructive measurements of quantum systems have been proposed~\cite{kuzmich1998atomic, kuzmich2000generation, scully1999quantum} and demonstrated~\cite{saffman2009spin, cox2016deterministic}, and have found applications in the fields of quantum simulation~\cite{georgescu2014quantum} and quantum metrology~\cite{giovannetti2006quantum, hosten2016measurement}. 
They have stimulated a new generation of quantum sensors including atomic clocks~\cite{ludlow2015optical, leroux2010orientation} and atom interferometers~\cite{tino2014atom, cox2016spatially}, which utilize the so-called spin-squeezed states~\cite{pezze2018quantum, ma2011quantum} that are capable of surpassing the standard quantum limit~\cite{itano1993quantum} given by the number of the atoms involved~\cite{kitagawa1993squeezed, wineland1994squeezed}. 
Such non-destructive measurements also assist in the realization of non-classical states of macroscopic systems~\cite{aspelmeyer2014cavity, chowdhury2020quantum} which can be used to probe quantum gravity effects~\cite{bonaldi2020probing}.
They also help pave the way for searches of new physics beyond the standard model~\cite{rosi2017quantum, obata2018optical}.

In a quantum system the value of a given variable can often be enclosed into a phase shift of light interacting with the observed system~\cite{pezze2018quantum}. It is often possible to arrange a situation where this phase shift is large for light only in a given frequency range~\cite{leroux2012unitary}.
Moreover, multiple interactions with the system as in an optical cavity~\cite{sorensen2003probabilistic}, can amplify this phase shift, reaching a metrological gain given by the collective cooperativity~\cite{tanji2011interaction} $N\eta$, where $N$ is the number of atoms and $\eta$ is the single-atom cooperativity,  which is proportional to the finesse of the cavity. 
However, there are many noise sources that can prevent a precise phase shift measurement with an optical cavity. Yet it is possible to arrange a differential measurement scheme, where the phase shift for the probe mode is large while for another reference mode it is negligible, allowing the common-mode cavity noise to be canceled. 

In this article, we report a phase shift measurement scheme with reduced cavity-length-induced phase noise using an optical ring cavity and two counter-propagating beams that function as probe and reference with a frequency difference of one cavity free spectral range (FSR). The proposed scheme has two advantages over the general noise cancellation scheme in a Fabry-P\'{e}rot cavity with single phase-modulated light~\cite{vallet2017noise, hobson2019cavity}. 
First, the ring cavity geometry allows for the manipulation~\cite{ling2001theory} and probing~\cite{salvi2018squeezing} of atomic momentum states as well as their internal states. 
Second, the scheme where two independent beams are simultaneously resonant with the cavity is very flexible in practical applications. 
The proposed system demonstrates close to $30$~dB reduction in the cavity length fluctuations down to the noise floor in a frequency range up to half the cavity linewidth ($\delta \nu/2 \simeq 30$~kHz). 
We further apply this measurement scheme in a simulated spin-squeezing experiment~\cite{salvi2018squeezing} where  a cavity phase shift measurement is performed with a $200~ \mu$s averaging time. We demonstrate an improvement in phase sensitivity by a factor of 40 with a phase resolution of $0.7$~mrad. 
With this improved phase resolution, the scheme removes one important barrier against attaining highly spin-squeezed states.

The article is organized as follows: section~\ref{sec:Theory} establishes the theoretical model for cavity noise cancellation; section~\ref{sec:Setup} describes the experimental setup; in section~\ref{sec:Results} the noise cancellation results are presented and the contributions from various noise sources are analyzed; in section~\ref{sec:Application} the noise cancellation scheme is applied in a simulated squeezing experiment and the potential improvement in squeezing is evaluated; finally, in section~\ref{sec:Conclusion} conclusions are given.

\section{Theoretical model} \label{sec:Theory}
\begin{figure}[h!]
   \centering
   \includegraphics[width=1\linewidth]{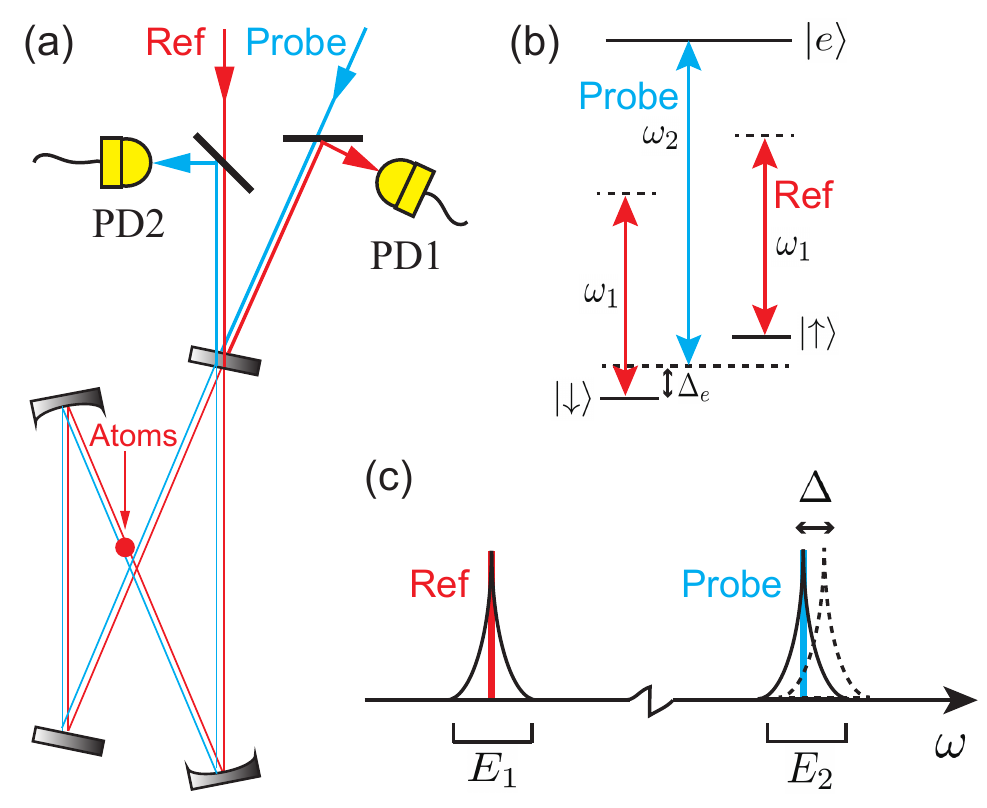}
   \caption{(a) Non-destructive phase shift measurement with reduced noise due to cavity length fluctuations. Laser cooled atoms (red circle) interact with the fundamental mode of an optical ring cavity and induce a shift $\Delta$ of the cavity resonance frequency. Two beams (Ref and Probe) are coupled to the cavity in counter-propagating directions and the reflections are collected by two photodetectors (PD1 and PD2).
(b) Simplified level diagram of the reference and probe beams with respect to the atomic transitions. The probe is close to the atomic resonance while the reference is far detuned. 
(c) Two beams are resonant with two modes of the cavity therefore the PDH error signals display common cavity-length fluctuations. In the differential scheme the atom-induced phase shift $\Delta$ can be resolved while the common-mode cavity noise can be suppressed.}
   \label{model}
\end{figure}

Even though the proposed cavity noise-reduced phase shift measurement scheme can be used in general quantum systems, here we focus on a particular application in a spin squeezing experiment~\cite{salvi2018squeezing},
where an optical ring cavity is used for the non-destructive measurement of the atomic momentum states (see Fig.~\ref{model}). In this proposal 20~dB squeezing is estimated considering only the atom shot noise versus the scattering into free space. 
In reality, the effect of cavity length fluctuations is not considered and might present a major obstacle. 
These cavity length fluctuations may originate from acoustic and sub-acoustic pressure changes, resonances of piezoelectric transducers (PZT) used to tune the cavity length, etc.
Taking into account the phase shift $\delta \phi$ induced by cavity length fluctuations, we express the cavity overall phase shift in the presence of the atoms as, 
\begin{equation}
\delta\Phi = 2 \Phi_{1} S_{z} + \delta\phi.
\label{Eq:overall_phase}
\end{equation}
With a detuning from atomic resonance $\Delta_{e}$ larger than the linewidth $\Gamma$ of the optical transition, $\Phi_{1} = \eta \Gamma/(2 \Delta_{e})$, where $\eta$ is the single-atom cooperativity and $S_{z} = (N_{\uparrow} - N_{\downarrow})/2$ is the atom number difference between two sublevels of the ground state (Fig.~\ref{model} (b)). At the atom shot noise limit, $S_{z}$ follows a Gaussian distribution with a standard deviation of $\sqrt{N}/2$.

We denote the atom-induced cavity phase shift as the \textit{signal} and the cavity-length-fluctuations-induced phase shift as \textit{noise} and compute the signal-to-noise ratio (SNR)  as, 
\begin{equation}
\mathrm{SNR} = \frac{(2\Phi_{1} \frac{\sqrt{N}}{2})^{2}}{\langle({\delta\phi})^{2}\rangle},
\label{gain}
\end{equation}
where the numerator is taken at the atom shot noise limit, $S_{z} = \sqrt{N}/2$ and $\langle \rangle$ denotes the expectation value. 
In order to resolve the atom-induced phase shift and achieve 20~dB squeezing, it is essential to suppress the cavity-length-fluctuations-induced phase noise down to a level 20~dB lower than the atom-induced phase shift.



The proposed noise-reduced phase shift measurement scheme is illustrated in Fig.~\ref{model}, where we consider two laser beams (Ref and Probe) with frequencies $\omega_1$ and $\omega_2$ that are resonant with two modes of an optical cavity at frequencies $\omega_{c1}$ and $\omega_{c2}$, respectively.
The resulting Pound-Drever-Hall (PDH) error signals~\cite{doi:10.1119/1.1286663}, $E_1$ and $E_2$, in the limit where the cavity resonance frequency fluctuations are small compared to the cavity linewidth, are proportional to the detunings $\delta_{c1}=\omega_1-\omega_{c1}$ and $\delta_{c2}=\omega_2-\omega_{c2}$. 
If the laser noise can be neglected, then $\delta_{c1}$ and $\delta_{c2}$ are proportional thus making it possible to consider a single detuning $\delta_c$ and a combination of $E_1$ and $E_2$ that is immune to cavity length fluctuations.
Taking into account additional, uncorrelated noise contributions to the error signals, $\delta E_1$ and $\delta E_2$, whose minimum variance is set by photon shot noise fluctuations, the two error signals can be expressed as,
\begin{equation}
E_1(t) = A_1 R_1(t)\ast \delta_{c}(t) + \delta E_1(t) ,
\end{equation}
\begin{equation}\label{eq:E2}
E_2(t) = A_2 (R_2(t)\ast \delta_{c}(t)-\Delta) + \delta E_2(t) ,
\end{equation}
where $A_{1}$ and $A_{2}$ are constants representing the amplitude of the signal.
In these expressions we have introduced the convolution with the response functions $R_1(t)$ and $R_2(t)$ which can arise from, e.g., electronic filtering, time delays or the response of the optical cavity. 
In this model, a constant shift $\Delta$ of the mode at frequency $\omega_{c2}$ is also introduced, as illustrated in Fig.~\ref{model} (c). This can be caused, for example, by the presence of a state-dependent index of refraction introduced by an atomic ensemble, as shown in Eq.~(\ref{Eq:overall_phase}). While temporal variations of $\Delta$ can be considered, here we assume that these are slow compared to the averaging time scale.
It is the main purpose of the proposed noise cancellation method to find a function $E$ of the error signals $E_1,E_2$ that maximizes the sensitivity to the shift $\Delta$.
To this end, we define the sensitivity error function $S$ as,
\begin{equation}\label{eq:sens_function}
S^2 = \frac{\mathrm{Var}(E)}{\left(\frac{\partial \langle E\rangle}{\partial \Delta}\big\rvert_{\Delta = 0}\right)^2},
\end{equation}
where $\mathrm{Var}$ denotes the variance.

It is instructive to first consider the trivial situation where $\delta E_1=\delta E_2 = 0$ and $R_1=R_2$. In this case one can see that $S^2$ is minimized and vanishes for a linear combination $E=E_1+\alpha E_2$ with $\alpha = -A_1/A_2$. If now the condition $\delta E_1=\delta E_2 = 0$ is relaxed, but the noise floor fluctuations remain small, i.e.  $\langle\delta E_i^2\rangle\ll A_i^2\langle\delta_c^2\rangle$, and $R_1 = R_2$, it is still possible to consider the linear combination $E = E_1+\alpha E_2$. In this limit, one can show that minimizing $S^2$ is equivalent to minimizing,
\begin{equation}
\mathrm{Var}(E) = \mathrm{Var}(E_1)+\alpha^2\mathrm{Var}(E_2)+2\alpha\mathrm{Cov}(E_1,E_2),
\label{alpha}
\end{equation}
where $\mathrm{Cov}$ denotes the covariance.
The minimum variance is attained when $\alpha = -\mathrm{Cov}(E_1,E_2)/\mathrm{Var}(E_2)$ and the resulting sensitivity error is,
\begin{equation}\label{eq:min_sens}
(S^2)_{\mathrm{min}}=\frac{\langle\delta E_1^2\rangle}{A_1^2}+\frac{\langle\delta E_2^2\rangle}{A_2^2} ,
\end{equation}
which is the sum of the noise floor contributions from the two error signals in frequency units.

We finally consider the case where the response functions $R_i$ differ. While determining the individual functions may not be experimentally straightforward, it is possible to measure the ratio of their Fourier transforms, i.e., the ratio of the  transfer functions $\tilde{R}=\tilde{R}_1/\tilde{R}_2$. Such a measurement can be performed, for example, by modulating the cavity length or the laser frequencies at a known frequency and then measuring the amplitude ratio and relative phase of the two error signals. 
Alternatively, in the presence of broadband cavity noise, as is our case, $\tilde{R}$ is determined by averaging the ratio of the Fourier transforms $\tilde{E}_1/\tilde{E}_2$, calculated from the (noisy) error signals. 
Once $\tilde{R}$ is determined, the noise cancellation can be applied to $E_1$ and the inverse transform of $\tilde{R}\tilde{E}_2=A_2\tilde{R}_1\tilde{\delta}_c + \tilde{R}\widetilde{\delta E}_2$. These two signals now share the same frequency response to cavity length fluctuations.

Finally, if $\tilde{R}$ differs from unity and one wishes to determine the residual cavity noise due to imperfect cancellation, it is first necessary to realize that the value of $\alpha$ determined as $-\mathrm{Cov}(E_1,E_2)/\mathrm{Var}(E_2)$ differs from the value $-A_1/A_2$ by a factor $\mathcal{I}=\int_0^{\infty}|\tilde{R}(\nu)|\cos(\phi_R(\nu)) \mathcal{S}_{\delta_c}^{(0)}(\nu)d\nu/\int_0^{\infty}\mathcal{S}_{\delta_c}^{(0)}(\nu)d\nu$, where $\tilde{R}=|\tilde{R}|e^{i\phi_R}$ and $\mathcal{S}_{\delta_c}^{(0)}=|\tilde{R}_2|^2\mathcal{S}_{\delta_c}$ is the spectral density of cavity frequency fluctuations multiplied by the amplitude of $\tilde{R}_2$. In this case, the residual cavity noise can be computed as,
\begin{equation}\label{eq:residual_PSD}
\delta S_E = A_1^2(|\tilde{R}|^2+\mathcal{I}^2-2\mathcal{I}|\tilde{R}|\cos\phi_R)S_{\delta_c}^{(0)}.
\end{equation}

\section{Experimental setup}\label{sec:Setup}

\begin{figure*}[t!]
   \centering
   \includegraphics[width=1\linewidth]{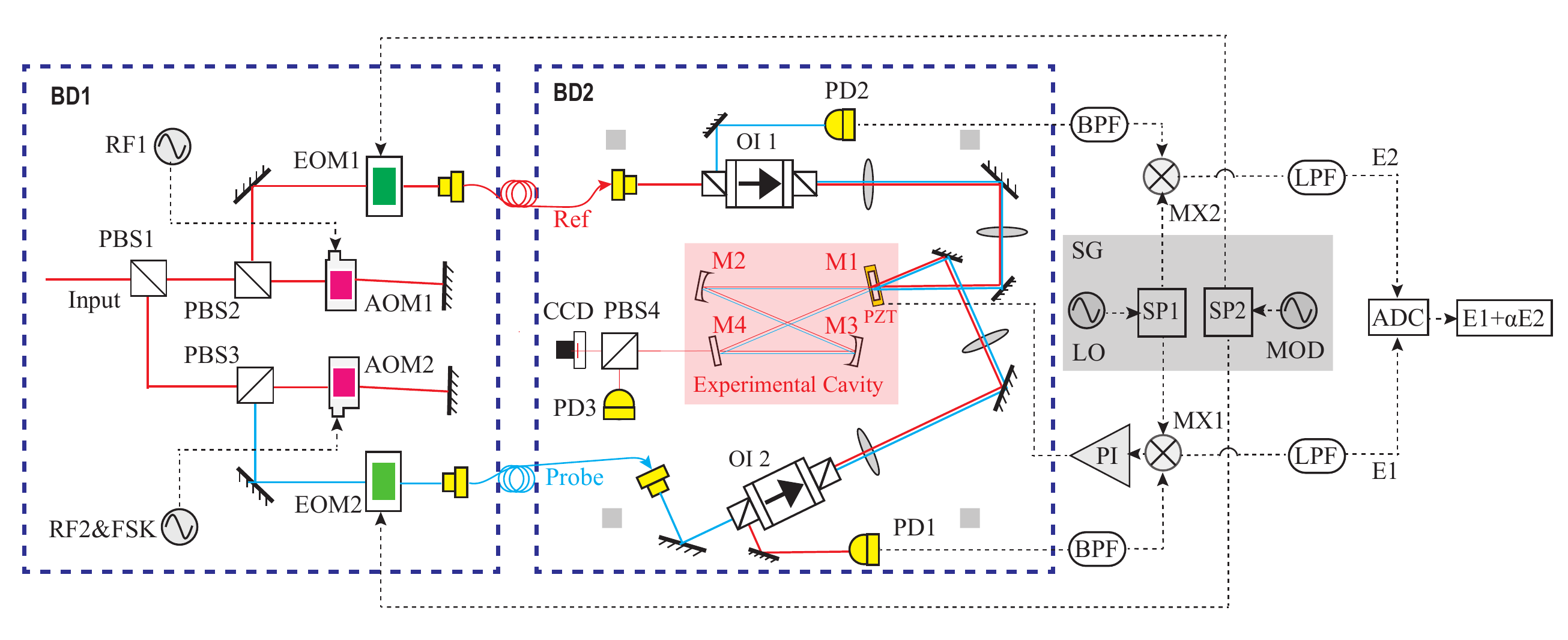}
   \caption{Experimental setup. BD1: two AOMs induce a frequency difference $\mathrm{FSR}  \simeq 1.43$~GHz between the two beams, indicated by the red and blue lines. AOM2 is driven by RF2, to which the FSK modulation can be applied. Two EOMs are driven at 10.5 MHz from the same source but with tunable relative phase and amplitude.
BD2: cavity, mode-matching optics and detection system. See text for details. 
Abbreviations: BD, breadboard; PBS, polarizing beamsplitter; AOM, acousto-optic modulator; EOM, electro-optic modulator; RF, radio frequency; FSK, frequency shift key; OI, optical isolator; PD, photodetector; CCD, charge-coupled device; SG, signal generator; LO, local oscillator; MOD, phase modulation; SP, splitter; MX, mixer; PI, proportional-integral controller; BPF, band-pass filter; LPF, low-pass filter; ADC, analog-to-digital converter; M, mirror; E, error signal. }
   \label{setup}
\end{figure*}

The core of the experimental setup is a bow-tie optical ring cavity with four high-reflectivity mirrors, shown in Fig.~\ref{setup}. The cavity mirrors are glued onto four V-shaped grooves with Torrseal epoxy and the grooves are glued on a stainless-steel cavity spacer with electrically conductive epoxy (EPO-TEK H20E), both epoxies are compatible with ultra-high vacuum. The V-groove that holds mirror~1 (M1 in Fig.~\ref{setup} ) is placed on a shear-force PZT (Noliac NAC2402-H2.3) in order to tune the cavity length. The whole cavity is assembled inside a glass box and is supported on sorbothane rubber balls for vibration isolation. Additionally, we can flow clean nitrogen through the box in order to reduce dust contamination. 

In Table~\ref{cavity_char} the main cavity properties are listed. The values of the mirror radius of curvature ($\mathrm{ROC}_{i}$) and of the mirror transmission $T_{i}$ are given in the sequence M1-M4, corresponding to the cavity scheme in Fig.~\ref{setup}. The mirror transmissions are specified for the laser wavelength of 689.448~nm, which corresponds to the $^{1}$S$_{0}$-$^{3}$P$_{1}$ intercombination transition of strontium (Sr) atoms.
The cavity FSR is measured by modulating the phase of  the input beam with an electro-optic modulator (EOM). The cavity transmission is increased when the frequency of the modulation matches the FSR, i.e., when the sidebands reach the adjacent cavity modes. This measurement yields an $\mathrm{FSR} = 1.43136(3)$~GHz. 
The cavity finesse is evaluated through the cavity ring-down method and the transmitted intensity decay fit yields an average photon lifetime of $\tau = 2.765(3)~\mu$s, or a linewidth of $\delta \nu = 1/(2\pi \tau) = 57.6(1)$~kHz and a finesse $\mathcal{F} = \mathrm{FSR}/(\delta \nu) = 2.40(2) \times 10^{4}$.

\begin{table}[h!]
\centering
\caption{\bf Relevant cavity parameters. }
\begin{tabular}{cccc}
\hline
Parameter &Symbol & Value & Units \\
\hline
Mirror ROC & $\mathrm{ROC}_{i}$ & $+\infty, 50, 50, +\infty$ & mm \\
Mirror transmission & $T_{i}$ & $250.8(3), <0.2, <0.2, 7.0(3)$ & ppm \\
Free spectral range & FSR & $1.43136(3)$ & GHz \\
Linewidth  & $\delta\nu$ &$57.6(1)$ & kHz\\
Finesse & $\mathcal{F}$ &$2.40(2)\times 10^{4}$ & $-$ \\
\hline
\end{tabular}
 \label{cavity_char}
\end{table}

The experimental setup for cavity noise cancellation and phase shift measurement  is also illustrated in Fig.~\ref{setup} and it is divided in two parts: (i) preparation of the two optical beams; (ii) measurement  and detection setup using the cavity. The two parts of the setup are placed on two independent breadboards, BD1 and BD2. While BD1 is fixed on the optical table, BD2 is placed on four pieces of sorbothane rubber for vibration isolation.  
The input laser light is frequency stabilized by locking to a high-finesse Fabry-P\'{e}rot cavity ($\mathcal{F}^{\prime} \simeq 8600$), reaching a 20~Hz laser linewidth~\cite{poli2006laser} and is transported to BD1. The input beam is split into two parts with a frequency difference of one FSR by two acousto-optic modulators (AOM) in double-pass configuration.
AOM1 is a high-frequency AOM (Brimrose) which shifts the frequency of the beam by $-1.21$~GHz, while AOM2 introduces a frequency shift of $+220$~MHz. 
The two beams are then phase modulated independently with two EOMs at 10.5~MHz and are transported to BD2 via two optical fibers. We refer to the two beams after the optical fibers as Ref and Probe, as shown in  Fig.~\ref{setup} and corresponding to the beams in Fig.~\ref{model}. 
On BD2, the two beams are independently mode-matched to the optical cavity. Optical isolators are used to couple two s-polarized beams to the optical cavity and detect the corresponding reflections from the back of the cavity incoupling mirror. The reflected beams emerging from the side ports of the optical isolators are detected via two homemade photodetectors (PD1 and PD2) in order to derive the two error signals using the PDH method.

PD1 and PD2 are low-noise and high-gain photodetectors with a bandwidth of 20~MHz, optimized for this application. The photodetector is based on a PIN photodiode (Hamamatsu S5821-01) and a transimpedance amplifier (TIA, OPA 657). The photodiode works in the reverse-biased mode and the TIA features a transimpedance gain of $100$~k$\Omega$. The simplified circuit schematic of the PD is illustrated in the inset of Fig.~\ref{PD_plot} in the Appendix. The PDs are powered by $12$~V batteries to eliminate noise from the mains electrical supply and the PD circuit is designed on a printed circuit board with surface-mount components. The output of the PDs are band-pass filtered at 10.5~MHz and are sent directly to a mixer (Minicircuits ZAD-1-1+) with no need for extra amplification.
A single two-channel signal generator is used to produce the local oscillator (LO) and modulation (MOD) signals required for both PDH signals. The two output channels are both split and sent to the independent mixers and phase modulators, respectively.
The output of the two mixers, which are the PDH error signals, are filtered by a second-order anti-aliasing low-pass filter (LPF) with a cut-off frequency of $f_{0} =  80$~kHz and a low-frequency gain of 10. 
This amplification reduces the relative contribution of the quantization noise of the analog-to-digital converter (ADC).
Finally, cavity locking is achieved by acting on the PZT under M1 using a standard PI controller with error signal $E_{1}$. Due to the limitation on the PZT response speed, the low frequency ($\lesssim 100$~Hz) noise is largely compensated by cavity locking, while the high frequency ($\gtrsim 100$~Hz) noise remains in the error signals and can be further suppressed by our noise cancellation scheme.



\section{Experimental results}\label{sec:Results}
In this section the results for the cancellation of the cavity length fluctuations are presented. With the proposed scheme the cavity length fluctuations can be canceled down to a level close to the noise floor, which is set by other noise sources. The contributions from various noise sources to the noise floor are also estimated.

\subsection{Noise cancellation performance}
In the following, the experimental sequence for data acquisition and the analysis are described.
We set the laser power of Ref and Probe (in Fig.~\ref{setup}) to be $40~\mu$W at the reflection which is detected by PD1 and PD2. When the cavity is scanning,  the error signals $E_{1}$ and $E_{2}$ exhibit a typical dispersive shape with a peak-to-peak voltage of $V_{pp}\simeq 2.45$~V.
When the cavity is locked with the PI controller acting on the PZT under M1, $E_{1}$ and $E_{2}$ show strong correlations since they both represent the cavity length fluctuations. The error signals are acquired by a digital oscilloscope for 10~ms with a sampling rate of 10~MHz. 
In order to analyze the data in the frequency domain,  we compute the Fast Fourier Transform (FFT) and estimate the voltage power spectral density (PSD) $\mathcal{S}_{V}(f)$ in a frequency range from 100~Hz to half the sampling rate, i.e., 5~MHz. 
The spectral density of frequency fluctuations can then be expressed as,
\begin{equation}
\mathcal{S}_{\nu}(f) = \left(\frac{\delta\nu}{V_{pp}} \right)^{2} \mathcal{S}_{V}(f) ~ ,
\label{PSD_Hz}
\end{equation}
where $\delta\nu = 57.6(1)$~kHz is the cavity linewidth.

\begin{figure}[h!]
   \centering
   \includegraphics[width=1\linewidth]{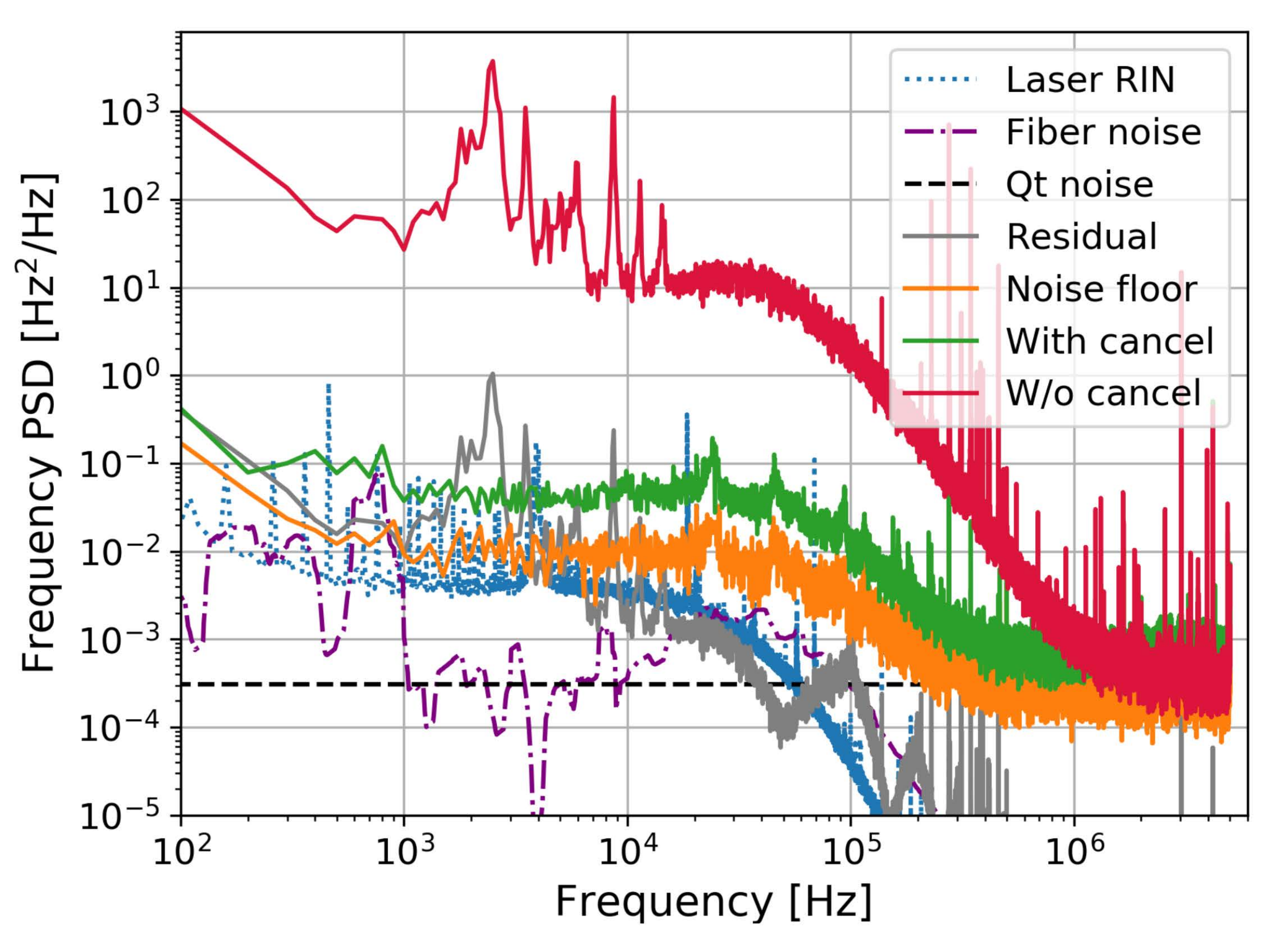}
   \caption{Cavity noise cancellation results. Red and green solid traces show the frequency PSD of the error signals before and after the noise cancellation, showing about $30$~dB noise reduction in a frequency range up to half the cavity linewidth ($\delta \nu/2 \simeq 30$~kHz). The noise of the canceled error signal is close to the noise floor (orange solid trace) determined when the laser is off-resonance with the cavity. Also shown are the residual noise due to the difference in the frequency response of the two channels (grey solid trace), the laser relative intensity noise (blue dotted trace), the quantization noise (black dashed trace) due to ADC and the fiber phase noise (purple dashdot trace).}
   \label{ALL_plot}
\end{figure}

The result of cavity noise cancellation is shown in Fig.~\ref{ALL_plot}. The red trace shows the original frequency fluctuations of $E_{1}$, while the green trace shows the dramatically reduced frequency fluctuations of the combined error signal $E = E_{1} + \alpha E_{2}$, where $\alpha=-\mathrm{Cov}(E_1,E_2)/\mathrm{Var}(E_2)$. 
The orange trace shows the noise floor corresponding to Eq.~(\ref{eq:min_sens}) and is obtained when the two laser beams are out of the cavity resonance.
Figure~\ref{ALL_plot} shows that the original cavity noise patterns are frequency-dependent. In the low-frequency range (100~Hz to 1~kHz), it follows a $1/f$ behavior indicating that flicker noise is dominating. In the mid-frequency range (1~kHz to 10~kHz), oscillations due to mechanical structures are dominating. For the cavity-aided phase shift measurement, we are interested in a bandwidth close to half the cavity linewidth ($\delta \nu/2 \simeq 30$~kHz). It is demonstrated that with our cavity noise cancellation scheme, within this frequency range the cavity noise can be reduced by more than 30~dB, close to the noise floor. 
At higher frequencies, up to 100~kHz, the cancellation scheme is still able to reach the noise floor, but the reduction is lessened due to the original cavity noise being strongly filtered.



\subsection{Noise sources analysis}
In the following we analyze the noise sources in our system and estimate their contributions to the noise floor. We investigate the effects from the laser intensity noise, the residual amplitude modulation (RAM) of the EOMs, the phase noise due to the fiber transportation, the different frequency responses of the two channels and the quantization noise in the ADC. The noise analysis of the PD is discussed in the Appendix.

\paragraph{Laser intensity noise} 

It is known that the PDH error signal is first-order immune to laser intensity fluctuations~\cite{doi:10.1119/1.1286663}.
In our system, however, the cavity is locked on one beam while the other beam can be tuned. If there is a mismatch between the laser frequency and the cavity resonance, then the laser intensity fluctuations may give a noise contribution in the PDH error signal. 
We cannot say a priori how large this frequency mismatch is, but we can estimate an upper limit to it. 
We observed that the amplitudes of the time domain error signals are within $1/5$ of the $V_{pp}$,
therefore we choose the upper limit of frequency mismatch as $1/5$ of the cavity linewidth $\delta \nu$. With this hypothesis we can estimate the maximum contribution of the laser intensity noise to the PDH error signal.

In order to measure the relative intensity noise (RIN), we illuminate the laser beam on PD1 and record the output for 10~s with an oscilloscope. We compute the PSD of this trace $\mathcal{S}_{V}^{rin}(f)$ and normalize it to the mean PD output voltage $V_{PD}$. Note that in the PDH method, the laser is filtered by the cavity and the error signals are filtered by a second-order LPF at the cut-off frequency of $f_{0} = 80$~kHz. Therefore, in order to compare the RIN with the noise floor in Fig.~\ref{ALL_plot}, the computed $\mathcal{S}_{V}^{rin}(f)$ should be corrected by the amplitude of the transfer function of the cavity, $|H_{cav}|^{2} = [1+(2f/\delta\nu)^{2}]^{-1}$, and that of the LPF, $|H_{LPF}|^{4} = [1+(f/f_{0})^2]^{-2}$.
Finally, the upper-limit contribution of the laser RIN to the PDH error signal in frequency PSD is,
\begin{equation}
	\mathcal{S}_{\nu}^{rin} (f) = \left(\frac{\delta\nu}{5}\right)^2\frac{\mathcal{S}_{V}^{rin}(f)}{V_{PD}^2} \cdot |H_{cav}|^{2} \cdot |{H_{LPF}}|^{4} . 
\end{equation}
The result is shown as the dotted blue line in Fig.~\ref{ALL_plot}, which has the largest contribution to the noise floor in a frequency range from 100~Hz to 10~kHz. 


\paragraph{Residual amplitude modulation (RAM)} 

The residual amplitude modulation arises from the imperfections in laser phase modulation when an EOM is used.  It has been studied extensively and has confirmed contributions from the etalon effect~\cite{whittaker1985residual}, the misalignment of light from the principal axis of the crystal~\cite{wong1985servo} and temperature variations, etc. Methods to actively cancel the RAM have also been demonstrated with a reduction down to the thermal noise level~\cite{zhang2014reduction}.  In order to estimate an upper limit of the noise contribution from the RAM,  we record the PDH error signals for 10~s when the laser is out of resonance with the cavity and compute the frequency PSD in a range from 100 mHz to 5 kHz. The results show that the noise contribution from the RAM of both the two EOMs are below $10^{-1}$ Hz$^{2}/$Hz at 10~Hz. At this level the RAM would not have an effect on the cavity noise cancellation since we are concerned about a frequency range where the contributions from the RAM are negligible. Indeed no active cancellation of the RAM is needed in our experiment.

\paragraph{Fiber phase noise}

As shown in Fig.~\ref{setup}, two 2-meter fibers are used for light transmission and mode-cleaning. Due to the pressure and temperature variations, the fiber transmission can introduce phase noise on the light, which can cause a phase shift in the cavity for the two beams and degrade the noise cancellation. To evaluate the differential phase noise introduced by the fiber transmission, 
we combined the two transmitted beams and measured the phase noise of the beatnote.
A fast photodetector is used to detect the 1.43~GHz beatnote and the output is sent to a phase noise analyzer (R\&S FSWP). Since the phase PSD $\mathcal{S}_{\phi}^{fiber}(f)$ is related to the frequency PSD by a factor of $f^{2}$, we can compute the frequency PSD due to the fiber phase noise as,
\begin{equation}
\mathcal{S}_{\nu}^{fiber}(f) = f^{2} \mathcal{S}_{\phi}^{fiber}(f) \cdot |H_{cav}|^{2}\cdot |{H_{LPF}}|^{4} , 
\end{equation}
where the transfer functions of the cavity response and the second-order LPF are considered.
The result is plotted as the purple dashdot trace in Fig.~\ref{ALL_plot}, which is well below the noise floor and has a negligible effect on the cavity noise cancellation. 

\paragraph{Frequency response difference between the two channels}
The difference in the frequency response of the two channels $E_1,E_2$ to cavity-length fluctuations may degrade the noise cancellation. However, as discussed in section~\ref{sec:Theory}, this difference can be compensated if it is a dominating noise source.
Different responses can originate from different polarizations of the two beams, accumulated phase shifts from electronics and optics, etc. We minimize this difference by using laser beams with the same polarization and cables with the same length for the RF signals. 
In order to measure the ratio $\tilde{R}=\tilde{R}_1/\tilde{R}_2$ of transfer functions for the two channels, we acquire 100 traces of $E_{1}$ and $E_{2}$ on resonance for 10~s with a sampling rate of 1~MHz. 
For each trace we compute the phase and amplitude of the ratio between the FFTs of the two channels and average over all the traces. We establish that the relative phase between the two channels is less than about 1\textdegree{}  in the relevant frequency range. We computed the residual cavity noise contribution by evaluating Eq.~(\ref{eq:residual_PSD}) and the result is shown as the grey solid trace in Fig.~\ref{ALL_plot}, thus showing that compensation of $\tilde{R}$ is unnecessary at the current level.

\paragraph{Quantization noise}

Quantization noise is introduced in the process of analog-to-digital conversion. In our data acquisition system,  a digital oscilloscope (Tek~MDO3014) is used to acquire the error signal data for 10~ms with a sampling frequency of $f_{s} = 10$~MHz. The 8-bit oscilloscope has a vertical resolution of $2^{8}-1 = 255$ and is set for a vertical full scale of  $\mathrm{FS} = 1$~V.  As a result, the least-significant-bit is $\mathrm{LSB} = \mathrm{FS}/(2^{8}-1) = 3.9$~mV and the 
one-sided voltage PSD is $\mathcal{S}_{V}^{qt}(f) = \mathrm{LSB}^{2}/(6f_{s})$. We compute the frequency PSD contribution due to quantization noise in the combined error signal $E$ as,
\begin{equation}
\mathcal{S}_{\nu}^{qt}(f) = \mathcal{S}_{V}^{qt}(f) \left(\frac{\delta \nu}{V_{pp}}\right)^{2} (1+\alpha^{2}) = 3.08\times 10^{-4}~ \mathrm{Hz}^{2}/\mathrm{Hz},
\end{equation}
where $\alpha = 1$ is used as an approximation.
The quantization noise is plotted as the dashed black trace in Fig.~\ref{ALL_plot}, it is clear that the quantization noise becomes dominant in the noise floor only in a frequency range higher than 100~kHz, which is beyond the cavity response and the effect can be neglected.

In summary, taking into account the PD noise analysis presented in the Appendix, we conclude that for the current setup, the main contributions to the noise floor come from the detection circuitry and the laser RIN, while the other noise sources and the effect due to different response between the two channels have negligible contributions. It is possible to further reduce the noise floor by adopting low-noise detection systems and by actively stabilizing the laser power.



\section{Application: Measuring a Cavity Phase Shift}\label{sec:Application}
\begin{figure}[t!]
   \centering
   \includegraphics[width=1\linewidth]{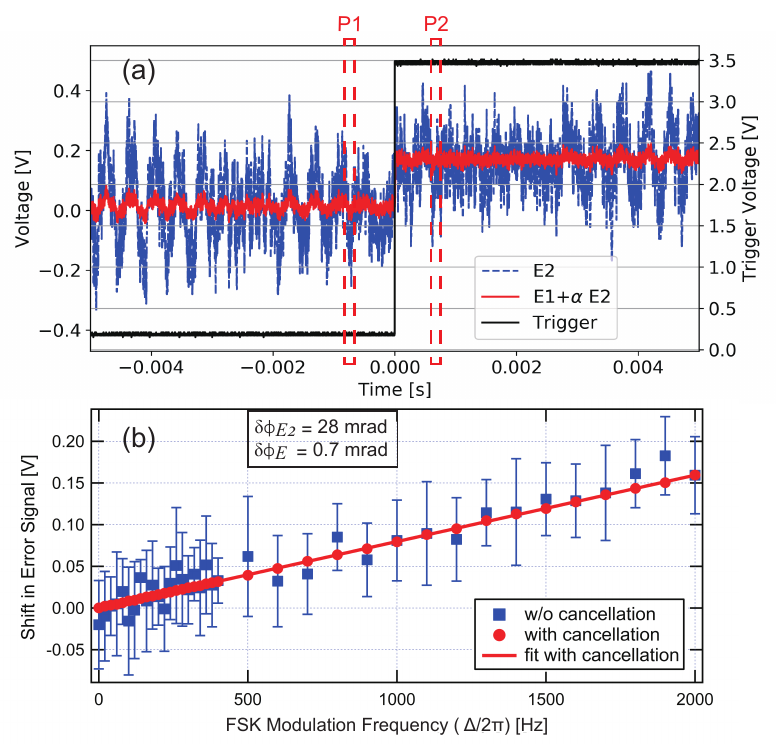}
   \caption{FSK modulation and sensitivity to laser frequency shift. (a) FSK modulation with 2~kHz laser frequency shift. The black trace is the trigger of the FSK, P1 and P2 are $T_{m} = 200 ~\mu$s-long probes with a delay time of 1~ms, representing the measurement sequence of a typical squeezing experiment. (b) A series of frequency shifts ranging from 20~Hz to 2~kHz is applied, each with ten acquisitions. The blue squares and red circles show the value of $\delta P$ for $E_{2}$ and $E$, respectively, error bars represent the standard deviation of 10 acquisitions. The red line is a linear fit of the red circle data. Inset shows the calculated phase resolutions  of $E_{2}$ and $E$, respectively.}
   \label{FSK_plot}
\end{figure}

The cavity noise cancellation method provides a powerful tool for the precise measurement of a phase shift of the light circulating inside the cavity.
We apply this scheme on a simulated squeezing measurement~\cite{salvi2018squeezing}, where we mimic the atom-induced cavity shift $\Delta$ in Eq.~(\ref{eq:E2}) by shifting the frequency of the Probe beam in Fig.~\ref{setup}. This frequency shift can be introduced through the frequency shift key (FSK) modulation on the RF source of AOM2.
Therefore, the Probe beam will be detuned from the cavity resonance by the amount of the FSK.
We apply an FSK modulation of 2~kHz and record the time domain traces of the error signal $E_{2}$ and compute the combined error signal $E$, as shown in Fig.~\ref{FSK_plot} (a). 
The black trace shows the trigger of the FSK modulation, where the data before the trigger are used for determining the value of $\alpha$ in Eq.~(\ref{alpha}) and the same value is applied on the whole dataset for noise cancellation.

As for a squeezing measurement, a typical probe time $T_m = 200 ~\mu$s is used and a differential scheme is adopted.
In our scenario, we simulate the squeezing measurement by extracting $200~ \mu$s data segments from both the non-shifted and shifted regions (P1 and P2 in Fig.~\ref{FSK_plot} (a)) with a delay time of 1~ms. We calculate the difference in the average of the two time series as $\delta P = \overline{P2} -\overline{P1}$ for both $E_{2}$ and $E$. 
For ten acquisitions with the same FSK frequency shift, the standard deviation in $\delta P$ for $E$ is reduced by a factor of 25 when compared to that of $E_{2}$, as shown by the error bars in Fig.~\ref{FSK_plot} (b).

For a more precise estimation of the measurement sensitivity to laser frequency (phase) shift, a series of FSK modulations from 20~Hz to 2~kHz is performed. Figure~\ref{FSK_plot} (b) shows the $\delta P$ values of $E$ (red circles) and $E_{2}$ (blue squares) as a function of the FSK modulation frequency, with error bars signifying the standard deviation of ten acquisitions.
We quantify the sensitivity to laser frequency shift with Eq.~(\ref{eq:sens_function}) by computing the frequency sensitivity $S = \sigma/a$, where $\sigma = \sqrt{\frac{n-1}{kn-1} \sum_{i=1}^{k} \sigma_{i}^{2}}$ is the weighted standard deviation of the error bars $\sigma_{i}$, $k = 37$ is the number of FSK frequencies, $n = 10$ is the number of acquisitions for each frequency and $a$ is the slope of the linear fit of $\delta P$ as a function of the FSK modulation frequency. 
For $E_{2}$ and $E$ we compute the frequency sensitivity as $S_{E_{2}} = 801$~Hz and $S_{E} = 20$~Hz, respectively, which can be converted into cavity phase resolution through $\delta \phi = S/(\delta \nu/2)$  as $\delta \phi_{E_{2}} = 28$~mrad and $\delta \phi_{E} = 0.7$~mrad, signifying an improvement in phase sensitivity by a factor of 40.
In order to prove the consistency of the frequency sensitivity measurements made with the FSK (Fig.~\ref{FSK_plot}(b)) and the measured frequency PSDs (Fig.~\ref{ALL_plot}), we evaluate the frequency sensitivity from the measured frequency PSDs, using the transfer function for the difference between averages of a time series. This yields a phase resolution of $\delta\phi_{E_{1}} = 24$~mrad and $\delta\phi_{E}^{\prime} = 0.5$~mrad, consistent with the results from the FSK measurement.

We can therefore use Eq.~(\ref{gain}) to estimate the SNR (in~dB) of the atom shot noise versus the cavity noise.  For realistic experimental parameters~\cite{salvi2018squeezing}, where $\eta \simeq 0.025$, $\Gamma \simeq 2\pi \times 7.6$~kHz is linewidth of the $^{1}$S$_{0}$-$^{3}$P$_{1}$ transition of Sr, $\Delta_e \simeq 2\pi \times 2.8$~MHz is the effective detuning from atomic resonance with electromagnetically induced transparency (EIT)~\cite{salvi2018squeezing} and $N \simeq 1\times 10^{5}$ atoms are involved, we estimate the atom-induced phase shift as $2\Phi_{1} \frac{\sqrt{N}}{2} \simeq 10.7$~mrad. 
Therefore the SNRs with and without the noise cancellation are $\mathrm{SNR}_{w} = 24$~dB and $\mathrm{SNR}_{w/o} = -8$~dB, respectively.
In the proposal paper~\cite{salvi2018squeezing}, 20~dB squeezing is estimated by considering only the atom shot noise versus the scattering into free space, the cancellation of the cavity noise to a level 24~dB lower than the atom shot noise makes the conclusion of this proposal solid, as the cavity noise would no longer play a dominant role. If instead the cancellation method were not applied, our current level of cavity-length fluctuations would completely mask the atomic signal.

\section{Prospect and conclusion}\label{sec:Conclusion}
In conclusion, we have demonstrated for the first time a phase shift measurement with pronounced immunity to cavity-length fluctuations using an optical ring cavity and two separate beams. 
We have achieved more than $30$~dB reduction in the cavity noise due to length fluctuations, close to the limit of the measured noise floor. We have applied this phase shift measurement scheme in a simulated spin squeezing experiment where we mimic the atom-induced cavity phase shift by changing the frequency of one of the two circulating laser beams. An improvement in phase sensitivity by a factor of 40 with a phase resolution of 0.7~mrad is achieved. With this method, squeezing up to 20~dB would not be limited by cavity-length fluctuations. 
This method is also applicable to two laser beams with largely different wavelengths as long as their frequency noise is negligible compared to the cavity resonance frequency fluctuations.


In the future, we will apply this method to quantum non-destructive measurements for the generation of spin-squeezed states in atom interferometers.
This method can find direct application to the cancellation of the effect of cavity length fluctuations in a cavity-aided non-destructive probe of Bloch oscillations~\cite{peden2009nondestructive} and Rabi oscillations~\cite{windpassinger2008nondestructive}.
More generally, it can assist in the non-destructive monitoring of quantum systems and find applications in the field of quantum simulation and quantum metrology.



\acknowledgements
We thank G. M. Tino and F. Marin for useful discussions and a critical reading of the manuscript. 
We acknowledge financial support from INFN and the Italian Ministry of Education, University and Research (MIUR) under the Progetto Premiale `Interferometro Atomico' and PRIN 2015.
JNT, LS and NP acknowledge support from European Research Council, Grant No. 772126 (TICTOCGRAV). 
EW acknowledges financial support from the program of China Scholarship Council (No.201703170201). 
GV acknowledges the receipt of a fellowship cofunded by the ICTP Programme for Training and Research in Italian Laboratories, Trieste, Italy and National Research Council, Italy.
We thank Mauro Chiarotti for his contribution in the initial stage of the experiment.

\appendix*
\section{Photodetector noise}
\label{PD_noise}
\begin{figure}[h!]
   \centering
   \includegraphics[width=1\linewidth]{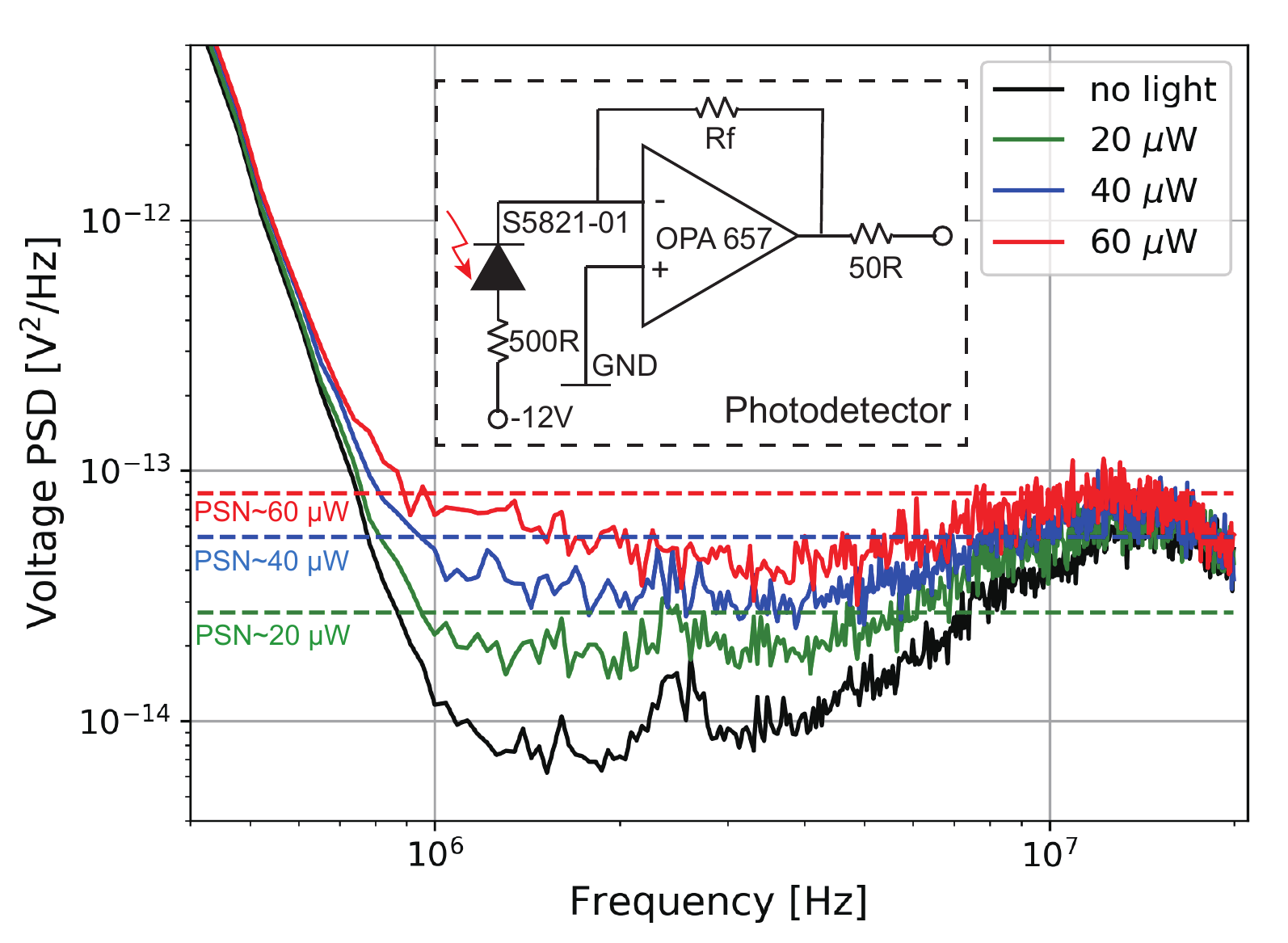}
   \caption{Photodetector noise measured with incident thermal light at different power levels and background noise floor without light. The PSD of PSN at different power levels is indicated by the dashed lines. Inset shows the simplified schematic of the photodetector, $R_f = 100$~k$\Omega$.}
   \label{PD_plot}
\end{figure}

We characterize the PD noise by illuminating the photodiode with thermal light, which is assumed to be photon shot noise limited. The output voltage of the PD is $V = R_{f} P s$, where $R_{f} = 100$~k$\Omega$ is the transimpedance gain, $s = 0.47$~A$/$W is the photon sensitivity of the S5821-01 photodiode, $P$ is the incident light power. We can therefore record the output voltage and convert it into light power.
The photodetector noise is measured with a spectrum analyzer with a resolution bandwidth of $100$ kHz, the converted voltage PSDs with different thermal light power as well as the background noise floor are shown in Fig.~\ref{PD_plot}, in a frequency range from 500~kHz to 20~MHz. 

We compare the photodetector noise with the photon shot noise (PSN) at the wavelength of $\lambda = 689$~nm. The PSD of the PSN is white and can be calculated as, 
\begin{equation}
\mathcal{S}_{V}^{psn} (f) = 2h\nu P {R_f}^2 s^2 ,
\end{equation}
where $h$ is Plank's constant and $\nu = c/\lambda$ is the laser frequency, $c$ is the speed of light in vacuum. 
The result is shown as the dashed lines in Fig.~\ref{PD_plot} where different colors represent different laser power levels. 

From the spectrum we can see that around the modulation frequency of $\simeq 10.5$~MHz, the PD is PSN limited at a power close to $40 ~\mu$W, which is typical for our measurement condition. This means that the PSN of the light is larger than the electronic noise of the PD. 
This response was measured also for the laser light and shows no significant difference at the modulation frequency compared to the thermal light.
The subsequent electronics, however, also contribute to the overall noise so that our PDH signal is not PSN limited for the given power level.

\bibliographystyle{apsrev4-1}
\bibliography{CavitypaperV19}

\end{document}